\documentclass[prl,twocolumn,aps,superscriptaddress]{revtex4}
\usepackage{amsmath}
\usepackage{graphicx}
\usepackage{tabularx}
\usepackage{amssymb}
\usepackage{epstopdf}
\usepackage{float}
\usepackage{color}

\begin{document}

\title{Vibrationally-induced electronic population inversion 
with strong femtosecond pulses}

\author{Pablo Sampedro}
\affiliation{Departamento de Qu\'imica F\'isica, Universidad Complutense, 28040 Madrid, Spain}
\email{pablosampedroruiz@gmail.com}

\author{Bo Y. Chang}
\affiliation{School of Chemistry (BK21), Seoul National University, Seoul 151-747, Republic of Korea}

\author{Ignacio R. Sola}
\affiliation{Departamento de Qu\'imica F\'isica, Universidad Complutense, 28040 Madrid, Spain}
\email{isola@quim.ucm.es}

\begin{abstract}
\noindent
We discover a new mechanism of electronic population inversion using
strong femtosecond pulses, where the transfer is mediated by vibrational
motion on a light-induced potential. The process can be achieved with
a single pulse tuning its frequency to the red of the Franck-Condon window.
We show the determinant role that the sign of the slope of the transition 
dipole moment can play
on the dynamics, and extend the method to multiphoton processes with
odd number of pulses. As an example, we show how the scheme can be applied
to population inversion in Na$_2$.
\end{abstract}

\maketitle


The development of femtosecond pulse technology has allowed the first 
experimental observation of transition structures that are fundamental 
to study chemical reactivity and kinetics\cite{Femto}; 
ultrafast imaging and structural dynamics are currently providing 
a wealth of spectroscopic and structural information concerning the 
dynamics of molecules in real time\cite{Imaging1,Imaging2}. 
Additionally, as the laser acts on the time scale of the
nuclear dynamics, ultrashort and strong pulses can be used not only to
monitor but also to control the movement of atoms, modifying
the yield and rate of chemical reactions\cite{QC1,QC2,QC4}. 
The use of closed-loop learning techniques has allowed a fast fully 
experimental approach to control quantum processes in molecules with strong
fields\cite{Rabitz1,Rabitz2}. 
However, the experimental observation alone is often
not enough to rationalize the outcome of the experiments. In
many cases, understanding the underlying processes requires the
use of models and theoretical simulations to describe the key
aspects of the dynamics. 
This is the approach followed in this work.

To maximize the transition probability between two quantum states coupled by an
external field is the fundamental challenge underlying most quantum control schemes.
When the interaction is coherent, it is possible to achieve full population 
inversion through Rabi flopping by controlling the pulse area\cite{Shore}. 
In molecules, however, 
the density of states forces the use of very long pulses, which make energy
relaxation and decoherence a major concern. Using shorter (and hence stronger)
pulses renders a new problem due to the Autler-Townes splitting generated by
the accessible yet unpopulated vibrational levels in the ground 
state\cite{Par1,Par3}.
This can be beaten by working with very short pulses, demanding even
stronger pulses\cite{Sampedro}.

Under strong fields, a plethora of new phenomena occur in molecules.
Aside from ionization or other multiphoton processes, the potential energy
surfaces change due to dynamic Stark shifts, dramatically affecting the
resonances and photophysics of the molecule\cite{SS1,SS2}.
One can modulate these laser-induced potential energy surfaces\cite{LIP3} (LIPs)
as drivers of the dynamics\cite{LIP4}.
Several photodissociation reactions were controlled in such way,
where under certain conditions the number of photons is conserved,
that is, the laser acts as a catalyst\cite{SF1,SF2,SF3}.


On the other hand, population inversion is typically controlled by chirping
the laser, with the frequency of the pulse sweeping across the absorption band, 
although by modulating the field the pulse duration is stretched\cite{ARP1,ARP2,ARP3,ARP4}.
It is possible however to achieve adiabatic passage with transformed-limited
pulses, as in the APLIP (adiabatic passage by light-induced potentials) 
scheme\cite{APLIP1,APLIP3}.
Several APLIP scenarios have been proposed\cite{APLIP2,APLIP4,APLIP4b}, 
but all of them require the use 
of at least two time-delayed pulses non-resonantly coupled to
an intermediate state. 
The key of the method lies in the presence of this intermediate electronic 
state that assists in modulating the LIP 
to guide the wave packet motion from one electronic state to the other.

If the pulses are long enough, this ``motion'' is in fact adiabatic or 
quasi-static, by which the wave packet always remains at the bottom of
the LIP preserving the initial vibrational quanta\cite{APLIP2,APLIP4b}. 
However, less-adiabatic or ultrafast APLIP are also possible\cite{Sampedro}.
In addition,
it has been proven that the APLIP principle can 
be extended to work for any system with an odd number of potentials ($N$), 
simply by using $N-1$ pulses\cite{APLIP7}.
While in this case the method maintains its robustness against changes in 
the field and its vibrational state selectivity, the intermediate electronic
states are populated transiently during the process, in contrast to APLIP.

In this work we want to demonstrate that a similar effect can be achieved 
using a single pulse that couples two electronic states slightly 
off-resonance.
In this case the electric field will induce a
different Stark effect on two different parts of the potential energy
surface, creating an effective LIP that will allow the system to oscillate
between the two coupled states. This asymmetry in the Stark effect can
be provoked by an antisymmetric (with respect to the internuclear distance)
 transition dipole too. 
However, the transition cannot be adiabatic in the nuclear motion. 
If the pulse is switched on slowly (its ramp up quite larger than the 
period of vibrational motion) then the wave packet will adiabatically
shift from the equilibrium geometry at $V_1$ to that at $V_2$, only
to revert the transition when the pulse is switched off.
The control mechanism that we propose in this work relies on the correlation 
between the vibrational motion and the electronic state for which we 
term the scheme the Vibrationally
Induced Electronic Transition in a Light-Induced Potential or VIETLIP.

\begin{figure}
\centering
  \includegraphics[width=0.9\columnwidth]{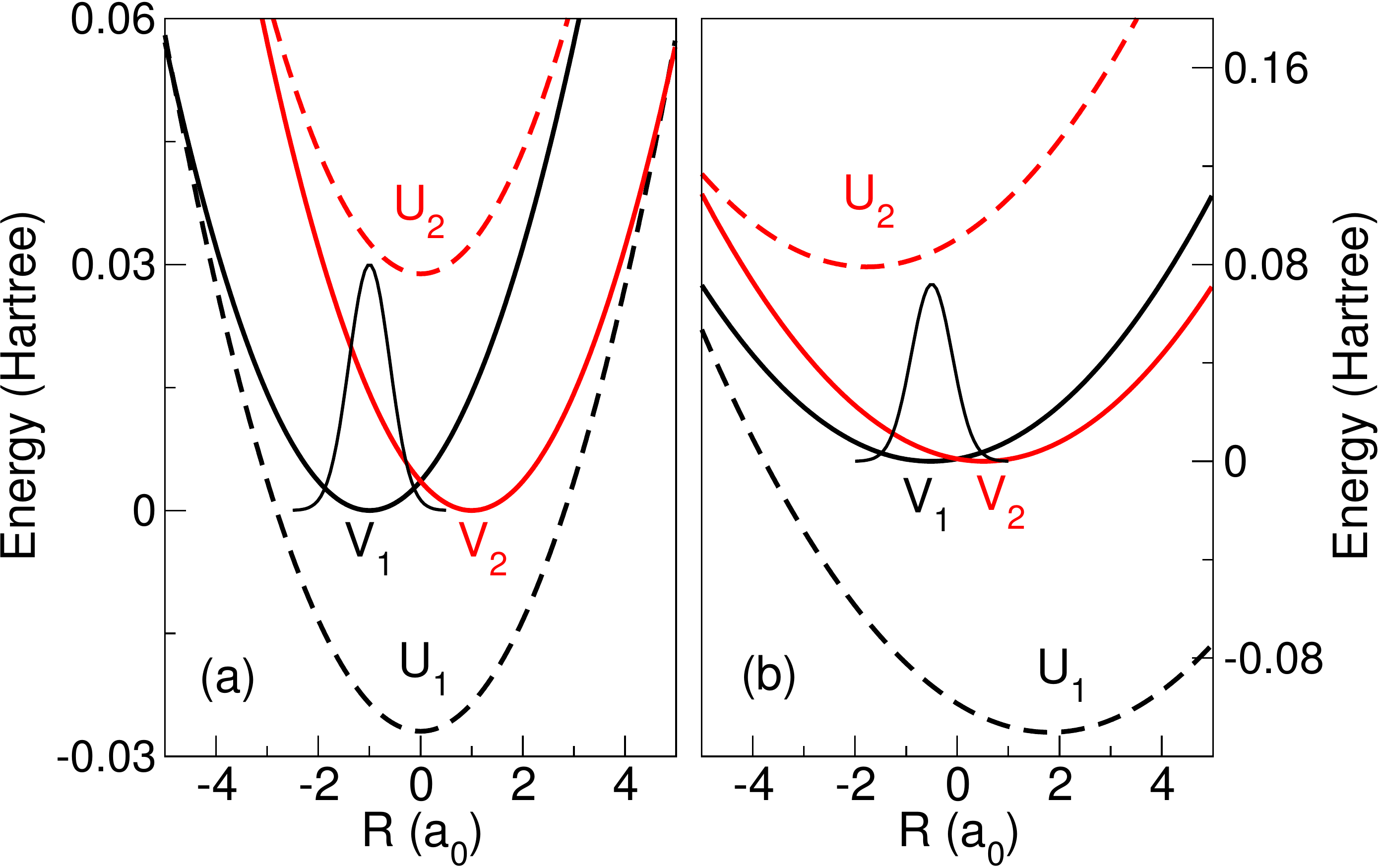}
\caption{Electronic potentials and LIPs generated by a strong field slightly
off-resonant from the absorption band. In (a) the equilibrium geometries
are more separated than in (b), so that the ground state wave function
(shown) overlaps excited state configurations in the latter case. In (b) the 
LIPs are calculated when the transition dipole depends linearly with $R$.}
\label{LIPtwoarmosc}
\end{figure}

Fig.\ref{LIPtwoarmosc} reveals such effect for the simplest system formed
by two harmonic oscillators, $V_1$ and $V_2$, coupled by a field. 
For illustration purposes we have chosen the reduced mass of the system to be 
that of Na$_2$ and the fundamental harmonic frequency $\omega$ to roughly
correspond to that of its ground electronic state.
The origin of the internuclear distance is chosen in between the equilibrium
geometries of both potentials.
In the first case [Fig.1(a)] we choose the excited state to be exactly as the 
ground state but shifted to a new equilibrium geometry, $R_1 - R_0 = \delta$, 
where $\delta = 2 a_0$ (approximately the displacement of the equilibrium
geometries of the ground and first excited electronic states of Na$_2$).
We assume a constant transition dipole.
Because we are not exciting at the Franck-Condon region, the transition
from $V_1$ to $V_2$ is hindered by an energy barrier $V_b$ that the
nuclear wave packet, initially in $V_1$, must overcome.

\begin{figure}
\centering
  \includegraphics[width=0.8\columnwidth]{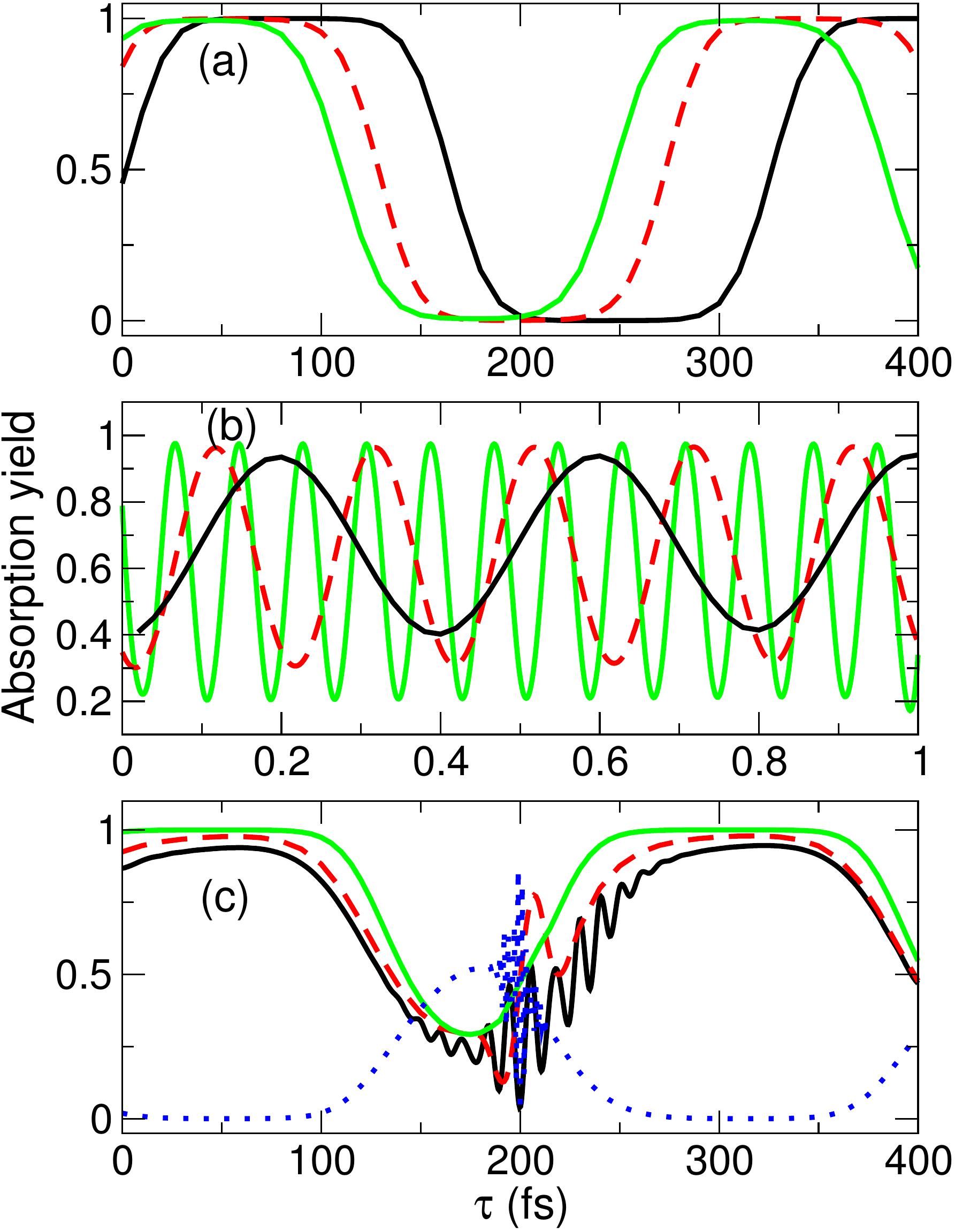}
\caption{Absorption yield as a function of the pulse duration (plateau, $\tau$)
using pulses of different peak amplitude: $0.1$ GV/cm (black), $0.21$ GV/cm 
(dashed red) and $0.43$ GV/cm (green). In (a) 
$\delta = 2 a_0$ while in (b) $\delta = a_0$. In both cases we assume a
constant dipole. In (c) $\delta = a_0$ but we assume a linear dipole.
In dotted lines we show the result when the dependence of the dipole
changes sign for the higher intense field.}
\label{TwoAOresonfreq}
\end{figure}

A strong field generates LIPs, $U_1$ and $U_2$: 
$U_1$ correlates with $V_1$ at short $R$ and with $V_2$ at large $R$.
Since it does not have any internal barrier, a nuclear wave packet initially
prepared in $U_1$ will freely move from $V_1$ to $V_2$. $U_2$ shows the
opposite correlation. 
Fig.2(a) shows the electronic population in $V_2$ at the end of the excitation
as a function of the pulse duration (plateau of the pulse), obtained after
integrating the time-dependent Schr\"odinger equation in a grid\cite{SO}
in the rotating wave approximation (RWA)\cite{Shore}.
We use plateau pulses with a relatively fast sine square switch on/off of
$60$ fs and a plateau of duration $\tau$. The electronic population follows
a slow squared-type oscillation with a period that depends on the period
of motion of the nuclear wave packet in the LIP. The period depends weakly on 
the pulse amplitude because the curvature of the LIP (and hence the harmonic
frequency) depends on the strength of the coupling. 
For weaker pulses the period will be larger.
In any case the period is of the order of hundreds of femtoseconds, an
order of magnitude (or more) larger than the Rabi period which depends on 
the pulse amplitude. For strong pulses the Rabi frequency is comparable to 
an electronic transition energy.

The VIETLIP scheme works as long as the equilibrium geometries of the 
electronic potentials $V_1$ and $V_2$ are separated enough, but if 
$\delta$ is too large, $E_b$ will be large too, and the scheme will
require very strong pulses to remove the energy barrier in the LIP.
On the other hand, if $\delta$ is smaller than the de Broglie wavelength
of the initial nuclear wave function, then part of this wave function
correlates with $U_1$ in the adiabatic representation (for $R< 0$), and
part of it with $U_2$ (for $R>0$), as shown in Fig.1(b). 
Under a strong field we generate a 
wave packet in two LIPs with no internal barriers, leading to 
interference and electronic beatings.
An example is shown in Fig.2(b) where we use the same potentials as before
but with $\delta = 1 a_0$. The population fully oscillates at a Rabi period
that depends exclusively on the Rabi frequency. 
For strong fields this implies electronic beatings in the order of the
femtosecond.

However, it is interesting to see that the presence of a coordinate-dependent
transition dipole can compensate this effect. 
In Fig.1(b) we show the LIPs when $\delta = 1 a_0$ but the dipole depends
linearly with the internuclear distance, $\mu = R$. The effect of the
dipole is to separate the equilibrium geometries of the LIPs, 
allowing to prepare the initial wave function in a single LIP.
The slope of the dipole (positive or negative) decides the 
the shape of the LIPs. If the coupling is $-R \times \epsilon$ 
(positive slope; larger dipole at large $R$)
where $\epsilon$ is the field, then, as in Fig.1(b) $U_1$ moves the wave 
packet towards $V_2$.
The dynamics is relatively similar (although
more complex) to that encountered in the first case. The oscillations
in the yield as a function of the pulse duration show a long period
corresponding to the nuclear motion, not to the Rabi oscillation.
However, if the coupling is $R \times \epsilon$ (negative slope; larger
dipole at small $R$) then
the equilibrium geometry of $U_1$ is at $V_1$. The wave packet is
relatively trapped at the ground electronic state, the trapping increasing
with the pulse intensity. The yield of absorption is therefore quite
smaller, as shown in Fig.2(c) and the absorption bands anti-correlate
with those obtained with opposite dipole.

\begin{figure}
\includegraphics[width=7cm]{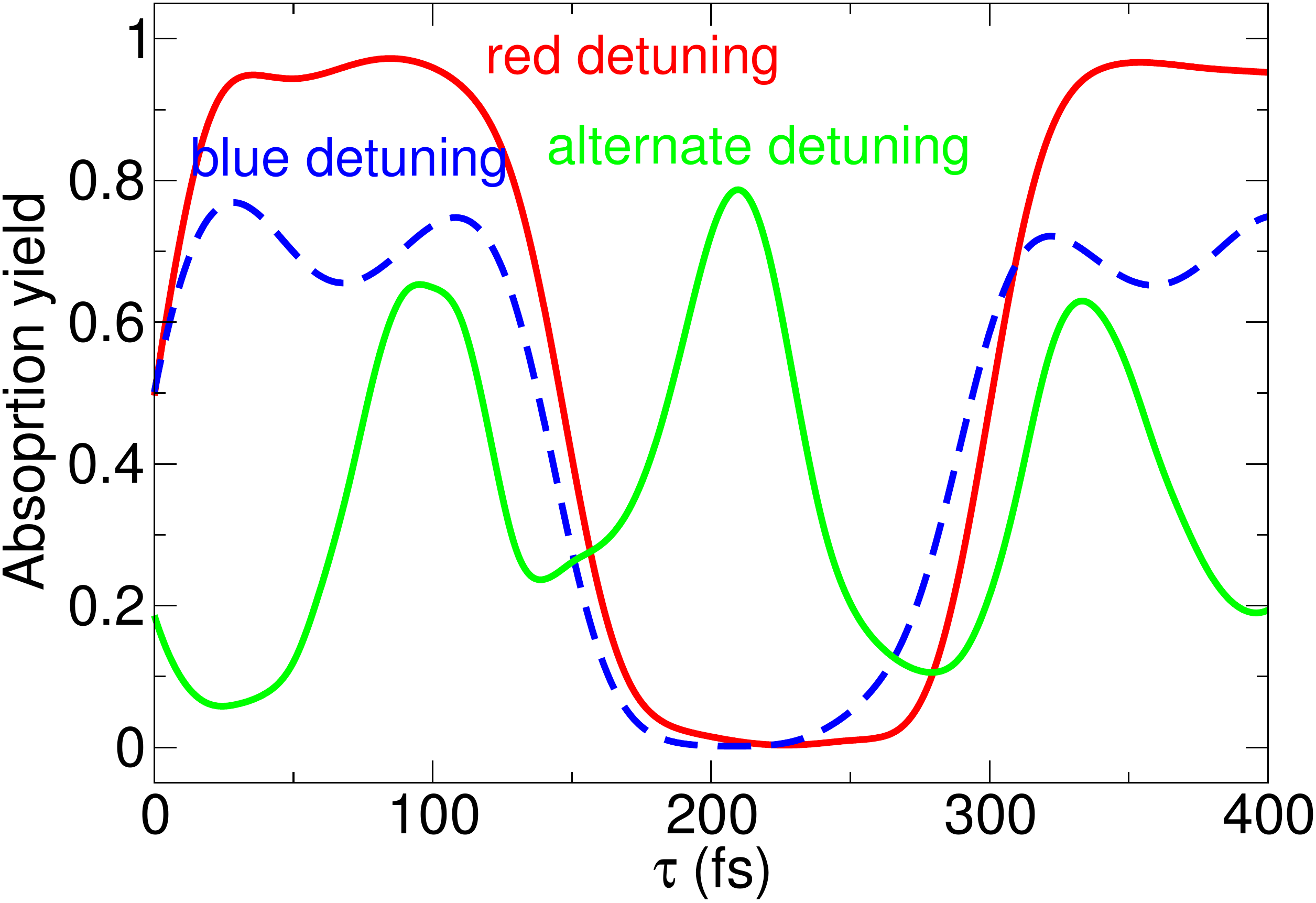}
\caption{Absorption yield as a function of the pulse duration (plateau, $\tau$)
for a $3$-photon process involving $3$ different pulses. The results depend
on the choice of the laser frequencies: the red-detuning configuration 
(intermediate potentials above $V_1$, $V_4$ or $\Delta > 0$), 
blue-detuning configuration ($\Delta < 0$) or alternate configuration
(one potential above and the other below the resonance.}
\end{figure}

One of the most interesting aspects of population transfer through diabatic
wave packet motion on a LIP is that the method can be extended to
multiphoton transitions, but only with an {\em odd} number of pulses,
contrary to the APLIP scheme, which works with an {\em even} number of pulses. 
Moreover the passage depends on the
choice of the laser frequencies: for some arrangements the passage is
more protected than for others.
In Fig.3 we show the results of population transfer between $4$ potentials,
$V_1$ to $V_4$, where we assume that there are only sequential couplings 
between nearest neighbors 
so that $V_1$ is coupled to $V_2$ by $\Omega_1(t)$, $V_2$ with $V_3$ by
$\Omega_2(t)$ and $V_3$ with $V_4$ by $\Omega_3(t)$.
As in multiphoton APLIP, the intermediate potentials must be off-resonance.
In our model we choose $V_2$ and $V_3$ as harmonic oscillators
centered 
at the equilibrium geometries of $V_1$ and $V_4$ respectively, 
although the results are not very sensitive to these parameters.
However, they are sensitive with respect to the choice of detuning:
$V_2$ and $V_3$ are displaced $\pm \Delta$ in the vertical axis. 
For the results in Fig.3 we
fixed the peak Rabi frequency as $0.05$ a.u. and $\Delta = 0.08$ a.u.
When $V_2$ and $V_3$ are above $V_1$ and $V_4$ (red detuning) full
population transfer by VIETLIP 
is possible. The results are worse in the blue-detuning configuration
($V_2$ and $V_3$ below $V_1$ and $V_2$) since the Stark effect
shifts the $V_1$ and $V_4$ potentials to higher energies.
Robust population transfer is still possible, particularly with stronger
fields.
However, in the alternate
configuration (when one potential lies above and the other below $V_1$ 
and $V_4$) the passage is clearly worse and less robust.
In principle, the same ideas can be generalized to any multiphoton transition
with odd number of pulses.



\begin{figure}  
\centering
\includegraphics[width=0.8\columnwidth]{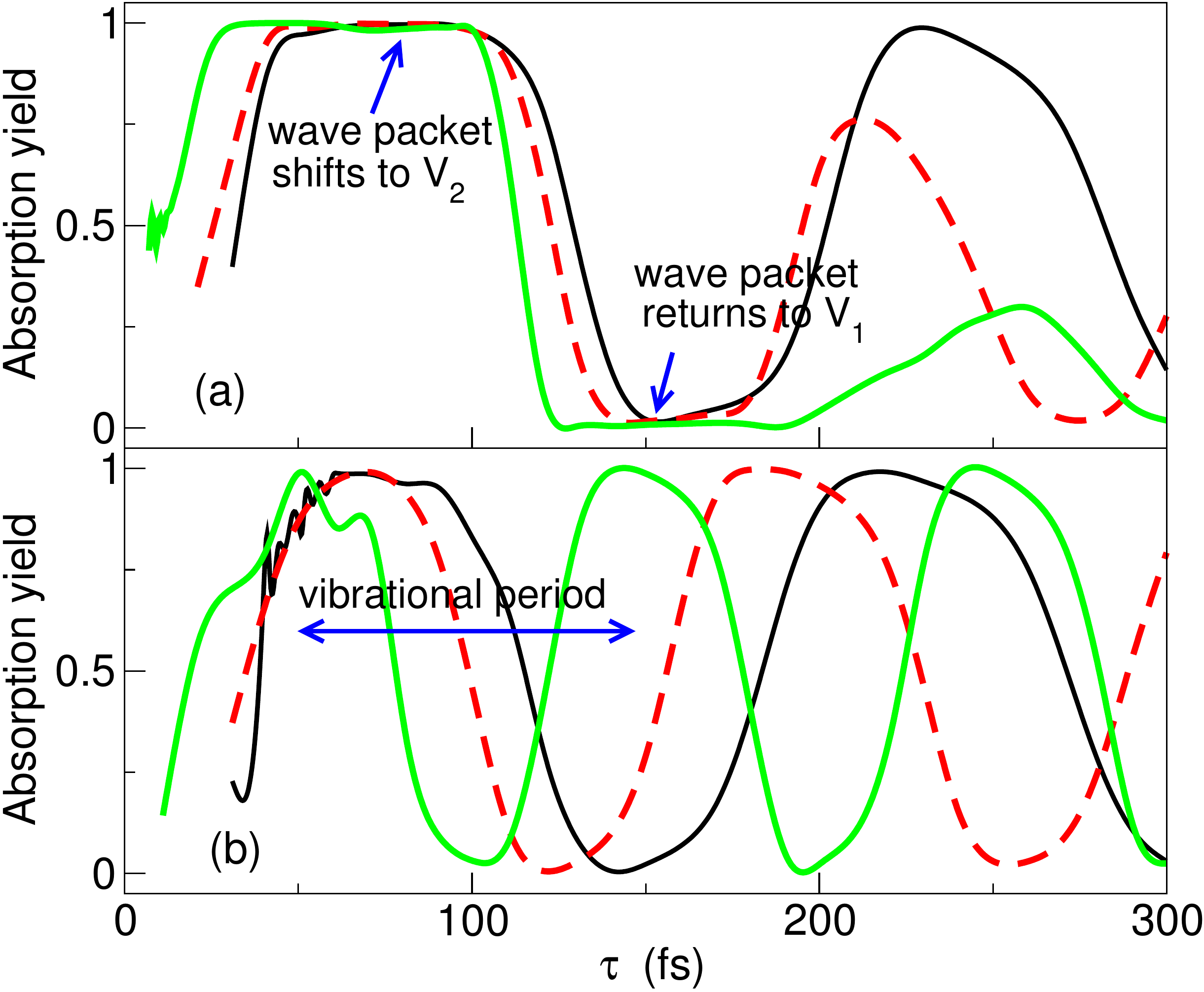}
\caption{Absorption yield for the $A$ band of Na$_2$ using $0.0027$ GV/cm 
(black), $0.054$ GV/cm (red, dashed) 
and $0.027$ GV/cm (green) 
peak amplitudes. The arrows mark the conditions where we show wave packet
dynamics in Fig.5. In (b) we assume a constant dipole.}
\label{CompNa2DM}
\end{figure}

Although the concept of LIPs is essential in understanding many processes
under strong fields, it has been experimentally difficult to show
evidence of wave packet transfer through LIPs. One of the main
difficulties for APLIP is to isolate the desired process from other competing
routes. As more pulses act on the system the RWA is typically violated
breaking the theoretical requirements for the transfer. Here we show
that the VIETLIP scheme can be applied in realistic conditions.
As an example, we show simulations for the dynamics in Na$_2$.
We use ab initio electronic potentials and dipoles for the transition
 between the ground state $^1\Sigma_g(3s)$ and the first excited state
$^1\Sigma_u (3p)$ that gives rise to the $A$ band.
Interestingly, the transition dipole is approximately linear in the
internuclear distance and has the appropriate sign. 
Choosing the frequency at the Franck-Condon window (with maximal
overlap with the vibrational levels in $A$) leads to saturation\cite{Par3}.
However, if we tune the laser below the resonance ($\omega = 1.8$ eV)
we obtain the results shown in Fig.4(a). 
Here we use Gaussian pulses of different peak amplitudes and pulse durations.
We observe square oscillations with long periods that
partially reproduce the behavior observed in Fig.2(a), although with 
quite lower pulse intensities (because of the large transition dipole).

\DeclareGraphicsExtensions{.pdf,.png}
\begin{figure}
\centering
\includegraphics[width=0.9\columnwidth]{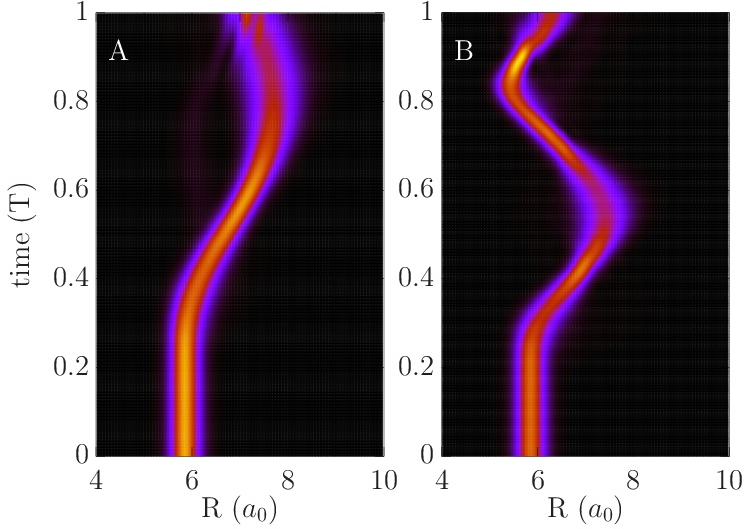}
 \caption{Wavepacket dynamics in Na$_2$ when the pulse duration leads to
population inversion (A) or transparency (B).}
  \label{WPdynamics}
\end{figure}

The dependence of the yield on the pulse duration again basically
follows from the vibrational motion of the wave packet in the LIP. 
Fig.5 shows the time evolution of the square of the wave packet for two
pulse durations: $75$ fs (within the maximum absorption band) and
$150$ fs (within the transparent conditions) using a pulse of
$2.7~10^{-3}$ GV/cm peak amplitude. If the wave packet has time to
move from $V_1$ to $V_2$ through the LIP (first oscillation) we have
maximum absorption. Doubling the time the wave packet returns to that
part of the LIP that correlates to $V_1$ as the pulse is switched off.

Comparing the yield of absorption in Na$_2$ with the model results
of Fig.2(a) shows some differences.
A striking difference is the decay of the second band of maximum absorption
(particularly with stronger fields) and the weak dependence of the position of
the bands on the pulse amplitude. This is an effect due to the dipole.
As observed in Fig.1(b) and Fig.2(c) the coordinate dependent dipole
deforms the LIPs affecting the diabatic motion of the wave packet in
the LIP, which is mostly trapped around the equilibrium geometry of the
LIP for large fields (because the $R \epsilon$ term increases with the
distance). For constant dipoles we fully recover the regular behavior,
as shown in Fig.4(b). The simulations in this case where performed
in the Na$_2$ potentials with an average constant dipole of $3.65$ ea$_0$.
Although the dynamics in the real Na$_2$ (with coordinate-dependent
dipole) are more complex, the dipole in fact makes population inversion
more robust, as evidenced by comparing the size of the absorption band
as a function of the pulse duration in both cases.

In summary, we have proposed a new robust scheme of population
inversion between two electronic states with displaced equilibrium geometries. 
The scheme shares many features with APLIP, as the mechanism
of population transfer is mediated by motion in a LIP. 
In VIETLIP, however, only one strong pulse is needed, that must be
tuned to the red of the absorption band, and shorter pulses can be used.
The process cannot be completed in a fully adiabatic way, so that the
pulse duration must be approximately synchronized to the vibrational
period.
Moreover the scheme can be extended to any multiphoton process
with odd number of pulses. 
Finally, we have found an intriguing dependence to the dipole function.
Obviously, a dipole going to zero for some internuclear distance
has always strong implications in adiabatic passage\cite{Photodiss}. 
In this work we report for the first time how the sign of its slope
determines the outcome of the transition, placing the molecular 
complexity in the forefront of the control process. 

\section*{Acknowledgments}
Financial support by the Spanish MICINN project CTQ2012-36184
and the Korean International cooperation program (NRF-2013K2A1A2054518) and 
Basic Science Research program (NRF-2013R1A1A2061898) is gratefully acknowledged.

\end{document}